# Quantum spin correlations through the superconducting-normal phase transition in electron-doped superconducting $Pr_{0.88}LaCe_{0.12}CuO_{4-\delta}$

Stephen D. Wilson[1], Shiliang Li,[1] Jun Zhao,[1] Gang Mu,[2] Hai-hu Wen,[2] Jeffrey W. Lynn,[3] Paul G. Freeman[4], Louis-Pierre Regnault[5], Klaus Habicht[6], Pengcheng Dai,[1,7,*]

[1]*Department of Physics and Astronomy, The University of Tennessee, Knoxville, Tennessee 37996-1200, USA*

[2]*National Laboratory for Superconductivity, Institute of Physics and National Laboratory for Condensed Matter Physics, Chinese Academy of Sciences, P.O. Box 603, Beijing 100080, China*

[3]*NIST Center for Neutron Research, National Institute of Standards and Technology, Gaithersburg, Maryland 20899-8562, USA*

[4]*Institut Laue-Langevin, 6, rue Jules Horowitz, BP156-38042 Grenoble Cedex 9, France*

[5]*CEA-Grenoble, DRFMC-SPSMS-MDN, 17 rue des Martyrs, 38054 Grenoble Cedex 9, France*

[6]*Hahn-Meitner-Institut, Glienicker Str 100, Berlin D-14109, Germany*

[7]*Neutron Scattering Sciences Division, Oak Ridge National Laboratory, Oak Ridge, Tennessee 37831-6393, USA*

[*]*To whom correspondence should be address. E-mail: daip@ornl.gov.*

**The quantum spin fluctuations of the S = 1/2 Cu ions are important in determining the physical properties of the high-transition temperature (high-$T_c$) copper oxide**

**superconductors, but their possible role in the electron pairing for superconductivity remains an open question. The principal feature of the spin fluctuations in optimally doped high-$T_c$ superconductors is a well defined magnetic resonance whose energy ($E_R$) tracks $T_c$ (as the composition is varied) and whose intensity develops like an order parameter in the superconducting state. We show that the suppression of superconductivity and its associated condensation energy by a magnetic field in the electron-doped high-$T_c$ superconductor, $Pr_{0.88}LaCe_{0.12}CuO_{4-\delta}$ ($T_c = 24$ K), is accompanied by the complete suppression of the resonance and the concomitant emergence of static antiferromagnetic (AF) order. Our results demonstrate that the resonance is intimately related to the superconducting condensation energy, and thus suggest that it plays a role in the electron pairing and superconductivity.**

**INTRODUCTION:** The parent compounds of the high-$T_c$ copper oxide superconductors are Mott insulators characterized by a very strong antiferromagnetic (AF) exchange in the $CuO_2$ planes and static long-range AF order. Doping holes or electrons into the $CuO_2$ planes suppresses the static AF order and induces a superconducting phase, with energetic short-range AF spin fluctuations that are peaked around the AF wave vector $\mathbf{Q} = (1/2, 1/2)$ in the reciprocal space of the two-dimensional $CuO_2$ planes (Fig. 1a)[1]. Understanding the relationship between the insulating AF and superconducting phases remains a key challenge in the search for a microscopic mechanism of high-$T_c$ superconductivity[2,3]. For optimally hole- and electron-doped high-$T_c$ superconductors, the most prominent new feature in the spin fluctuation spectrum is a collective magnetic excitation known as the resonance mode, which is also centered at **Q** = (1/2, 1/2) and whose characteristic energy ($E_R$) is proportional to $T_c$[4-8]. The resonance only appears below the superconducting transition temperature in these optimally-doped systems and is fundamentally linked to the superconducting phase itself.

The resonance previously has been suggested as contributing a major part of the superconducting condensation[9], measuring directly the condensation fraction[10], and possessing enough magnetic exchange energy to provide the driving force for high-$T_c$ superconductivity[11-13], but its small spectral weight compared to spin-waves in the AF insulating phase may disqualify the mode from these proposed roles[14]. One way to determine the microscopic origin of the resonance is to test its relationship to the superconducting condensation energy. Strictly speaking, the notion of superconducting condensation energy is an ill-defined concept if the normal state fluctuation effects are important as in the case of hole-doped high-$T_c$ copper oxides[15,16]. However, in the absence of an accepted microscopic theory, one may still use the mean-field expression to estimate the condensation energy in order to determine if the mode can indeed contribute to the interaction necessary for electron pairing and superconductivity[14]. Within the *t-J* model, a direct determination of the magnetic exchange energy available to the superconducting condensation energy requires the knowledge of the wave vector and energy dependence of the *normal* state spin excitations at zero temperature[17]—a quantity that has not been possible to obtain due to the presence of superconductivity. In principle, this can be rectified by studying the evolution of the zero (low) temperature spin excitations through the superconducting-to-normal state phase transition using magnetic field as a tuning parameter. Unfortunately, the large upper critical fields ($H_{c2} >$ 30 T) required to completely suppress superconductivity in optimally hole-doped superconductors prohibit the use of neutron scattering in such a determination. In the lower field measurements on $La_{2-x}Sr_xCuO_4$ (LSCO), neutron scattering experiments have found that a magnetic field causes intensity to shift into the zero-field spin gap at the expense of the resonance[18, 19]. This is consistent with the idea that the resonance is being gradually pushed into the elastic channel where a quantum critical point (QCP) separates the superconducting state from an AF state[20, 21]. Raman scattering results, however, showed that the primary effect of an applied field is simply to increase the volume fraction of the AF phase at the expense of the superconducting phase, thus suggesting an intrinsic electronic phase separation of these two phases[22].

Electron-doped superconductors require a much lower upper critical field ($H_{c2}$ < 10 T) to completely suppress superconductivity[23], thereby enabling one to probe the evolution of the spin excitations, resonance, and static AF order in these materials as the system is transformed from the superconducting into the normal state at low temperature. Here we present electronic specific heat, elastic and inelastic neutron scattering results on the optimally electron-doped superconductor $Pr_{0.88}LaCe_{0.12}CuO_{4-\delta}$ (PLCCO, $T_c$ = 24 K)[8]. We show that a magnetic field that suppresses the superconducting condensation energy in PLCCO also suppresses the resonance in a remarkably similar way (Fig. 1d). Furthermore, the reduction in magnetic scattering at the resonance energy with increasing magnetic field is compensated by the intensity gain of the elastic scattering at the AF ordering wave vector $\mathbf{Q}$ = (1/2, 1/2, 0) (Figs. 1e and 1f). Therefore, the superconducting phase without static AF order can be directly transformed into an ordered AF phase without superconductivity in electron-doped PLCCO via the application of a magnetic field. These results present the possibility that the resonance is intimately related to the electron pairing and superconductivity.

**RESULTS AND DISCUSSION:** We used inelastic neutron scattering experiments on the IN-8, IN22, BT-9, and V2 triple-axis spectrometers to map out the field dependence of the magnetic scattering function, $S(\mathbf{Q}, \omega)$, over a range of energies ($0 \leq \hbar\omega \leq 18$ meV) in electron-doped PLCCO. We chose to study PLCCO because the crystalline electric field (CEF) ground state of $Pr^{3+}$ in PLCCO is a nonmagnetic singlet and $Ce^{4+}$ is nonmagnetic[24], thus greatly simplifying the interpretation of the data. Additionally, as will be discussed later, nearly optimally doped PLCCO ($T_c = 24$ K) has an experimentally determined and easily accessible upper critical field, $H_{c2} = 7$ T (Fig. 1b), necessary for the complete suppression of the superconducting phase[25].

Since previous work on hole-doped superconducting $YBa_2Cu_3O_{6.6}$ showed that a moderate $c$-axis aligned magnetic field can suppress the intensity of the resonance[12], we first probed the influence of such a field on the recently discovered resonance in electron-doped PLCCO (ref. 8). Figures 2a and 2c show **Q**-scans through (1/2, 1/2, 0) at the resonance energy ($E_R \approx \hbar\omega = 10$ meV) in zero field on the IN-8 and BT-9 triple-axis spectrometers, respectively. Consistent with earlier observation[8] and the new polarized neutron beam measurements (see Supporting Information), cooling from the normal ($T = T_c + 6$ K) to the superconducting ($T \approx T_c - 20$ K) state clearly enhances the magnetic scattering at (1/2, 1/2, 0), which we define as the resonance (Fig. 1e). After applying a field greater than ($H = 10$ T) or near ($H = 7$ T) $H_{c2}$, these same **Q**-scans show that the superconductivity-induced enhancement (the resonance) in zero field (Figs. 2a and 2c) has now been completely suppressed, leading only to normal state AF spin fluctuations (Figs. 2b and 2d). Since the $Pr^{3+}$ CEF excitations are weakly wave vector dependent[24], the magnetic field-induced suppression at $\mathbf{Q} = (1/2, 1/2, 0)$ must arise from the reduction of $Cu^{2+}$ spin fluctuations at the resonance energy.

Figures 3a-e summarize the energy dependence of the scattering obtained on IN-8 at the peak center [$\mathbf{Q} = (1/2, 1/2, 0)$] and background [$\mathbf{Q} = (0.6, 0.6, 0)$] positions (Figs. 1a and 2b) for temperatures below and above $T_c$ (*i.e.* 5 K and 30 K) and under zero, 8 T, and 10 T fields. Turning first to data collected using a neutron final energy of $E_f = 14.7$ meV (Fig. 3a), the results at zero field are consistent with earlier measurements (see Fig. 3 in Ref. 8) and show enhanced magnetic scattering around 10 meV below $T_c$ indicative of the resonance. Our polarized neutron beam measurements confirmed the magnetic nature of the mode (see Supporting Information). A $Pr^{3+}$ CEF excitation at $\hbar\omega = 6$ meV is also observed in both signal and background scans[8]. While application of an 8 T magnetic field has no influence on the nonmagnetic background and the 6 meV CEF excitation at 5 K, there is a clear suppression of scattering at the resonance energy of 10 meV at 5 K (Fig. 3a). To cover a wider energy range around 10 meV at $\mathbf{Q} = (1/2, 1/2, 0)$, we carried out measurements using $E_f = 35.0$ meV (Figs. 3b and 3c). In the zero field case, the temperature difference data (Fig. 3d) again show a clear resonance peak at 11 meV, identical to the results of polarized neutron beam measurements (see Supporting Information). However, this well-defined resonance peak vanishes when a 10 T field is applied, as confirmed by comparing the energy-integrated (from 8 to 16 meV), superconductivity-induced intensity changes between zero ($314 \pm 77$ counts/10 mins, Fig. 3d) and 10 T ($-19 \pm 85$ counts/10 mins, Fig. 3e). Future experiments at lower fields are necessary to determine if the suppression observed in $\mathbf{Q}$-scans of Figs. 2a-d is due to a downward shifting of the mode's energy with increasing field, as seen in hole-doped LSCO (refs. 18,19).

In order to probe the field-suppressed spectral weight distribution of the resonance, we have carried out systematic elastic scattering measurements across $\mathbf{Q} = $ (1/2, 1/2, 0) under various magnetic fields. While previous muon spin relaxation (μSR) measurements have observed the emergence of field-induced AF order throughout the volume of optimally-doped PLCCO (ref. 26), neutron diffraction experiments have failed, possibly due to insufficient sample volume, to detect such a signal for optimally and over doped PLCCO (refs. 27, 28). Since our PLCCO samples have no observable static AF order above $T = 0.6$ K (ref. 8), we would expect the observed peak at (1/2, 1/2, 0) in Fig. 4b to be nonmagnetic and thus weakly temperature dependent. This is indeed the case as shown in scattering profiles between 2 K and 30 K (Fig. 4b). However, when an 8 T $c$-axis aligned magnetic field is applied, field-induced magnetic intensity appears at the $\mathbf{Q} = $ (1/2, 1/2, 0) position (Fig. 4a). The field subtraction (8 T – 0 T) data at $T = 2$ K in Fig. 4c show this induced order without the presence of the residual nonmagnetic structural peak at (1/2, 1/2, 0). The temperature dependence of the scattering reveals that the AF field-induced order increases with decreasing temperature (Fig. 4g), remarkably similar to the field-induced incommensurate elastic scattering from hole-doped $La_{1.9}Sr_{0.1}CuO_4$ [29] and $La_2CuO_{4+y}$ [30].

To demonstrate that the field-induced effect observed in Fig. 4c is indeed related to the suppression of superconductivity, we note that $H_{c2}$'s in copper oxides are highly anisotropic with respect to the direction of an applied field. While a $c$-axis aligned field can suppress superconductivity most efficiently, the same field parallel to the $CuO_2$ planes would have a substantially reduced effect on superconductivity. On the contrary however, magnetic signal resulting from a simple polarization of paramagnetic $Pr^{3+}$ moments in PLCCO is smaller for fields along the *c*-axis compared to that in the *ab*-plane (see Supporting Information). Figure 4d shows **Q**-scans through (1/2, 1/2, 0) on an identical PLCCO ($T_c$ = 24 K) sample aligned in the [*h, h, l*] zone such that the vertical field ($H \approx 6.8$ T) was applied along the in-plane wave vector [-1, 1, 0] direction. The field subtraction results (Fig. 4e) indicate that there is no observable induced magnetic signal at 2 K. Now, re-aligning the *same* crystal in the [*h, k, 0*] zone on the *same* spectrometer with the *same* magnet, the field subtraction results reveal a clear induced peak at (1/2, 1/2, 0) (Fig. 4f). This is direct and unambiguous evidence that the field effect in PLCCO is associated with the suppression of superconductivity.

Figure 4h shows that the induced-order increases approximately linearly with increasing field up to $H_{c2}$. This is remarkably similar to the induced static moment seen in $La_{1.856}Sr_{0.144}CuO_4$ (ref. 31), which is thought to be near a QCP between the "mixed-phase" (where superconducting and AF phases coexist) and phase-pure superconducting phase[20, 21]. Since both hole-doped $La_{1.856}Sr_{0.144}CuO_4$ (ref. 31) and electron-doped PLCCO ($T_c$ = 24K) [8] have no static AF order at zero field, field-induced AF order cannot be due to modifications of residual AF order. With the present data, however, it is difficult to decipher any lower critical field threshold necessary for the emergence of static AF order in this PLCCO sample. Future experiments are needed to precisely map out the detailed field dependence of this field-induced AF order in PLCCO ($T_c$ = 24 K), which would in turn allow a more complete assessment of the concomitant suppression of the resonance mode and the creation of static AF order under field.

To compare neutron measurements with the superconducting heat capacity anomaly, a small piece cut from one of the crystals[8] studied in our neutron measurements was used to measure the electronic specific heat under various field strengths (Figs. 1b and 1c). Similar to previous work on optimally electron-doped $Pr_{1.85}Ce_{0.15}CuO_4$ (ref. 25), the entropy in PLCCO is almost conserved between the normal and superconducting states for $0 < T < T_c$ (see Supporting Information). This suggests that the fluctuation effects crucial for obtaining the correct superconducting condensation energy in hole-doped materials[15,16] are much less important for optimally electron-doped cuprates[25]. Using mean-field theory, we estimate the superconducting condensation energy for PLCCO, along with the upper critical field ($H_{c2}$) necessary for the complete suppression of the superconductivity [$H_{c2}(T = 0) \sim 7$ T], and the results are plotted in Fig. 1d. The physical quantity referred to here as the condensation energy is calculated in terms of the entropy loss measured at a given $T$ and field strength $H$ through the relation

$$U_c(T) = \int_T^{T_c+15K} [S_N(T') - S_{SC}(T')]dT'.$$

Using the specific heat determined upper critical field and the data from Figs. 2 and 3, we plot schematically the behavior of $S(Q, \omega)$ at $\mathbf{Q} = (1/2, 1/2, 0)$ in the fully superconducting ($H = 0$) versus the superconductivity-suppressed ($H > H_{c2}$) states in Figures 1e,f. The resonance is only observed in the superconducting state, while it disappears at high field, where the spectral weight losses at the resonance and quasi-elastic energies (See Supporting Information) are compensated in part by the intensity gain at the elastic AF position (Fig. 1f). This is different from the case of hole-doped LSCO (refs. 18,19) and electron-doped $Nd_{1.85}Ce_{0.15}CuO_4$ (ref. 32).

Since the reduction of the resonance intensity with increasing field in PLCCO parallels the suppression of the superconducting condensation energy (Fig. 1d), it is tempting to think that magnetic excitations contribute a major part of the superconducting heat capacity anomaly and condensation energy (Figs. 1b -1d). For optimally hole-doped YBa$_2$Cu$_3$O$_{6.95}$, the change in the magnetic excitations between the normal and superconducting states was $\langle m^2 \rangle_{res} = 0.08 \pm 0.014$ $\mu_B^2$/Cu and the condensation energy was $U_c = 1.5$ K/Cu, thus giving a ratio $\langle m^2 \rangle_{res} / U_c = 0.06 \pm 0.009$ $\mu_B^2$/K (ref. 13). In PLCCO, the integrated moment of the resonance is a much weaker $\langle m^2 \rangle_{res} = 0.0035 \pm 0.0014$ $\mu_B^2$/Cu (see Supporting Information); however, the condensation energy also has a much smaller value of $U_c = 0.0687$ K/Cu (Fig. 1c), rendering a similar ratio of $\langle m^2 \rangle_{res} / U_c = 0.05 \pm 0.02$ $\mu_B^2$/K. While this estimation in itself does not prove that magnetic excitations contribute a major part of the condensation energy, it is clear that the resonance is intimately related to the electron pairing and superconductivity.

The surprising observation of a simple trade off in intensities with increasing field between the resonance associated with the superconducting phase and the AF order in the nonsuperconducting phase is consistent with Raman scattering results[22]. These results suggest that the AF and superconducting phases compete with each other. However, it is unclear whether the AF ordered phase is associated with vortices[33] and therefore microscopically phase separated from the superconducting phase, or is uniformly distributed throughout the sample as suggested by μSR measurements[26]. The remarkable parallel between the suppression of the resonance and condensation energy with increasing magnetic field also suggests that the mode is fundamentally connected to superconductivity and the entropy loss associated with the phase's formation. Finally, our experiments elucidate a direct transition from a pure superconducting state without residual static AF order to an AF ordered state without superconductivity. Such a transition is not expected in conventional superconductors, and therefore can be used to test theories for high-$T_c$ superconductors[20, 21, 33, 34, 35]. Future absolute measurements of magnetic excitations over a wider energy and momentum space in the low-temperature superconducting and non-superconducting normal states should enable a more quantitative determination of the magnetic exchange energy contribution to the superconducting condensation energy, and thus help identify the driving force for electron pairing and high-$T_c$ superconductivity.

**MATERIALS AND METHODS:** Our inelastic neutron scattering experiments on electron-doped PLCCO ($a = b = 3.98$ Å, $c = 12.27$ Å; space group: *I4/mmm*) were performed at the IN-8, IN22, and BT-9 thermal triple-axis spectrometers at the Institute Laue Langevin and the NIST Center for Neutron Research, respectively. Cold neutron data were collected on the V2 triple-axis spectrometer at the Hahn Meitner Institute. Here we denote positions in momentum space using $\mathbf{Q} = (h, k, l)$ in reciprocal lattice units (rlu) in which $\mathbf{Q}$ [Å$^{-1}$] = ($h\, 2\pi/a$, $k\, 2\pi/b$, $l\, 2\pi/c$). The applied magnetic field was vertical, and the copper oxygen layers of the compound were aligned either in the horizontal scattering plane or perpendicular to it.

**ACKNOWLEDGEMENTS:** We thank Eugene Demler, Hong Ding, and Ziqiang Wang for helpful discussions. The neutron scattering part of this work is supported in part by the U.S. NSF DMR-0453804. The PLCCO single crystal growth at UT is supported by the U.S. DOE BES under grant No. DE-FG02-05ER46202. ORNL is supported by the U.S. DOE under contract No. DE-AC05-00OR22725 through UT/Battelle LLC. The work at IOP, CAS is supported by NSFC, the CAS project ITSNEM and the MOST project (2006CB601000, and 2006CB92180).

**Figure 1 Specific heat measurements of the superconducting condensation energy and summary of neutron scattering results for PLCCO ($T_c$=24K). a**, *Upper panel:* The two-dimensional $CuO_2$ plane. *Lower panel:* Schematic of typical constant-energy scans through reciprocal space. Spin excitations are centered at **Q** = (1/2, 1/2, 0). **b**, Field dependence of the total electronic specific heat versus temperature. Data taken at 8 T were established to be above $H_{c2}$ (ref. [25]) and were used to isolate and subtract background contributions from the normal state phonon/electronic heat capacity. To obtain the normal state electronic specific heat $\gamma T$, 8 T data are fitted by $C = \gamma T + \beta T^3$, where $\beta T^3$ is the phonon contribution. The resulting linear electronic contribution $\gamma T$ ($\gamma$ = 5) was added back to the field subtracted data to obtain the total electronic specific heat. **c**, Field subtracted measurements of the specific heat $(C_{SC} - C_N)/T$, versus temperature. The resulting entropy loss $S_N(T) - S_{SC}(T) = \int_0^T (C_N - C_{SC}) dT'/T'$ can then be calculated. **d**, Condensation energy, $U_c$, determined from $U_c(T) = \int_T^{T_c+15K} [S_N(T') - S_{SC}(T')] dT'$ and plotted as solid blue square symbols connected by a solid line. Intensity of the resonance mode plotted as a function of applied field. Red triangles denote peak intensity measurements at **Q** = (1/2,

1/2, 0), $\hbar\omega$ = 11 meV at $T$ = 4 K with the normal state background at $T$ = 30 K subtracted. For field strengths greater than 6 T, entire **Q**-scans were performed in order to resolve the resonance excitation. The 5 K – 30 K peak intensities of Q-scans at 6.8 T and 10 T taken from Gaussian fits on a linear background (whose raw data are shown in Figs. 2a-d) are plotted as open green (6.8 T) and teal circles (10 T). **e**, Schematic plots of the zero-field $S(\mathbf{Q}, \omega)$ at **Q** = (1/2, 1/2, 0) below and above $T_c$. **f**, For $H > H_{c2}$, the complete suppression of the resonance mode is observed along with the simultaneous appearance of a static AF order.

**Figure 2 Inelastic neutron measurements showing the suppression of the resonance mode under a c-axis aligned magnetic field.** **a**, Zero field **Q**-scans at $\hbar\omega$= 10 meV at 5 K and 30 K on IN-8 with 60'-60'-S-60'-open collimations with neutron final energy fixed at $E_f$ = 14.7 meV. The spectral weight increase below $T_c$ demonstrates the presence of the resonance mode. **b**, 10 T **Q**-scans at $\hbar\omega$ = 10 meV again on IN-8 showing no difference between 5 K and 30 K intensities. Throughout our experiments, magnetic fields are always applied above 30 K and the samples were field-cooled to low temperature. **c**, **Q**-scans on BT-9 with 40'-48'-S-40'-80' collimations showing the resonance intensity again at 10 meV in 0 T using $E_f$ = 28 meV. **d**, Identical Q-scans at 5 K and 30 K showing the disappearance of the resonance mode under 6.8 T.

**Figure 3 Inelastic neutron measurements of magnetic field effect on the energy dependence of the spin fluctuations in PLCCO ($T_c$=24K). a**, Energy-

scans at (1/2,1/2, 0) taken on IN-8 at both 5 K and 30 K in 0 T and 8 T. The data show a clear suppression in 5 K scattering intensity at the resonance energy, $E_R$, under 8 T field. Background points were collected at **Q** = (0.6, 0.6, 0). The 6 meV peak originates from a CEF excitation (ref. 8). **b**, Energy scans at 5 K and 30 K at (1/2, 1/2, 0) in 0 T on IN-8. Background data at (0.6, 0.6, 0) and 30 K are also plotted. **c**, Energy scans at 5 K and 30 K at (1/2, 1/2, 0) in 10 T on IN-8. Background data were collected at (0.6, 0.6, 0) at both 5 K and 30 K. **d,e**, Temperature subtracted (5 K – 30 K) energy scans collected at (1/2, 1/2, 0) under 0 T and 10 T, respectively. The energy integrated intensity from 8 to 16 meV is $314 \pm 77$ counts/10 mins at 0 T and $-19 \pm 85$ counts/10 mins at 10 T.

**Figure 4 Elastic neutron data demonstrating field-induced antiferromagnetic order under a c-axis aligned magnetic field in PLCCO ($T_c$=24K).** **a**, Elastic **Q**-scans through (1/2, 1/2, 0) in 0 T and 8 T at 2 K. Fits to the data are Gaussian line-shapes on a linear background. Data were collected on V2 using 60'-open-S-open-open collimations and $E_f$ = 5 meV with a cold Be filter before the analyzer. **b**, Elastic **Q**-scans under 0 T at 2 K and 30 K. **c**, $T$ = 2 K field (8 T- 0 T) subtraction of data shown originally in **a**. **d**, Low-$T$ elastic **Q**-scans through (1/2, 1/2, 0) under 0 T and 6.8 T fields along the [-1, 1, 0] direction (in the CuO$_2$ planes). Data were collected on BT-9 with 40'-48'-S-40'-80' collimations and 3 pyrolytic graphite filters. **e**, Field subtraction (6.8 T – 0 T) data with $H \parallel$ [-1, 1, 0] as shown in **d**. **f**, Identical field subtraction (6.8 T – 0 T) data with $H \parallel$ [0, 0, 1]. **g**, Temperature dependence of elastic intensity under

both 0 T and 7 T. $T_c$ is denoted by the vertical dashed line, while a fitted constant value for the 0 T intensity is shown as a dashed horizontal line. **h**, Field dependence of peak intensity values measured at 2 K and **Q** = (1/2, 1/2, 0), $\hbar\omega$ = 0 meV with 0 T, 30 K measured background value subtracted. The peak intensity value obtained via a Gaussian fit to the 8 T data shown in **c** is plotted as well. The solid line is a linear fit.

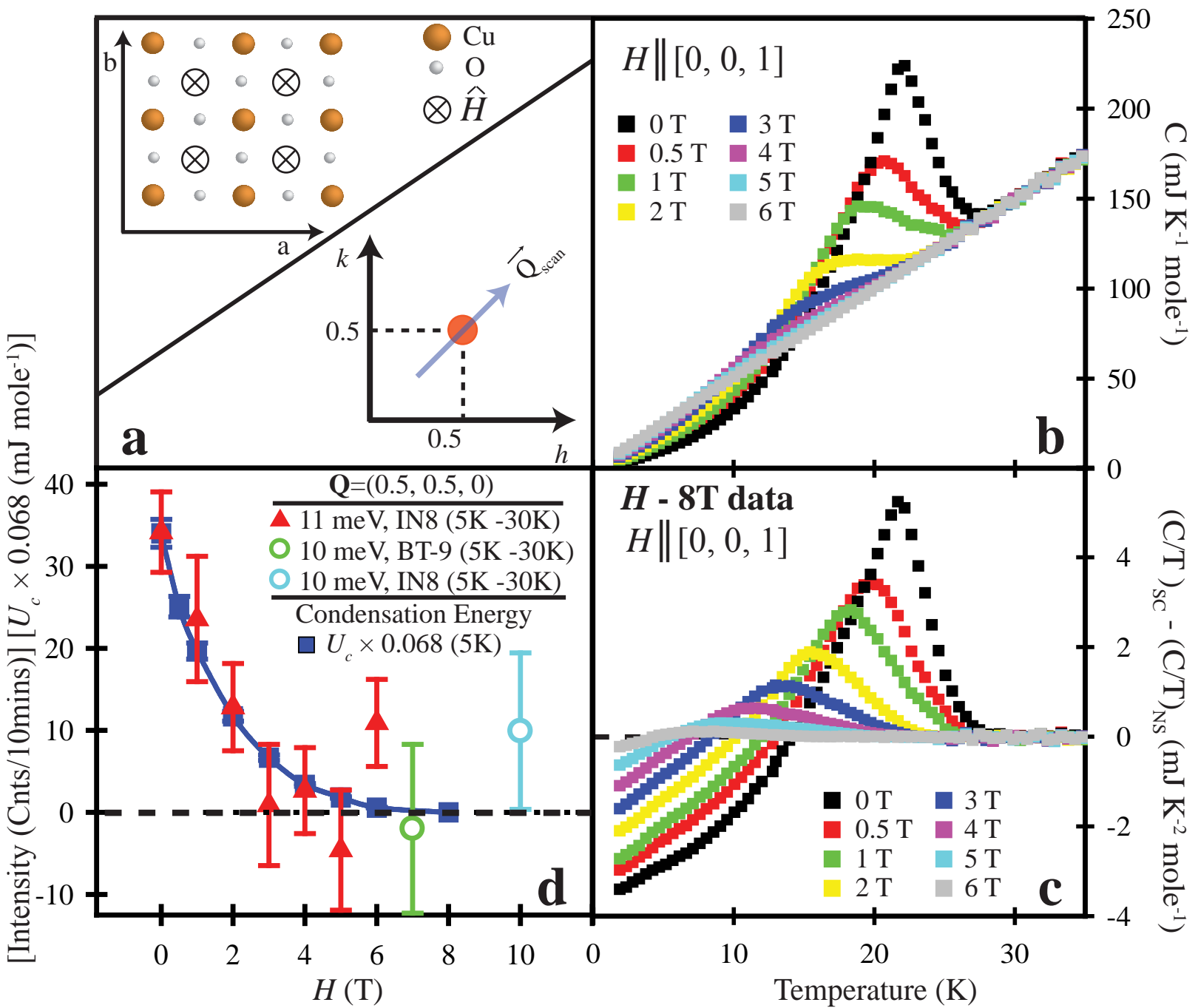

Figure 1

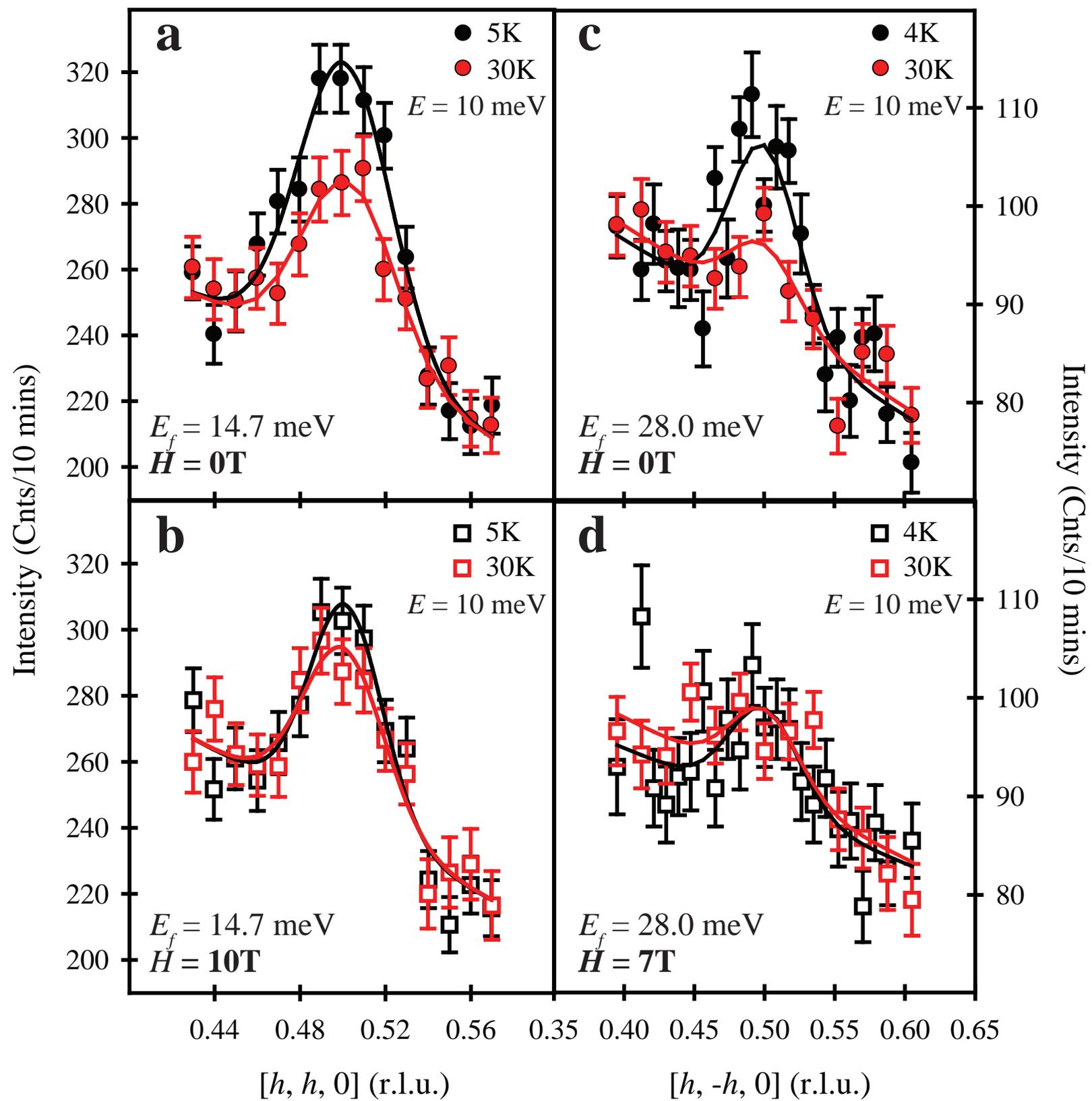

Figure 2

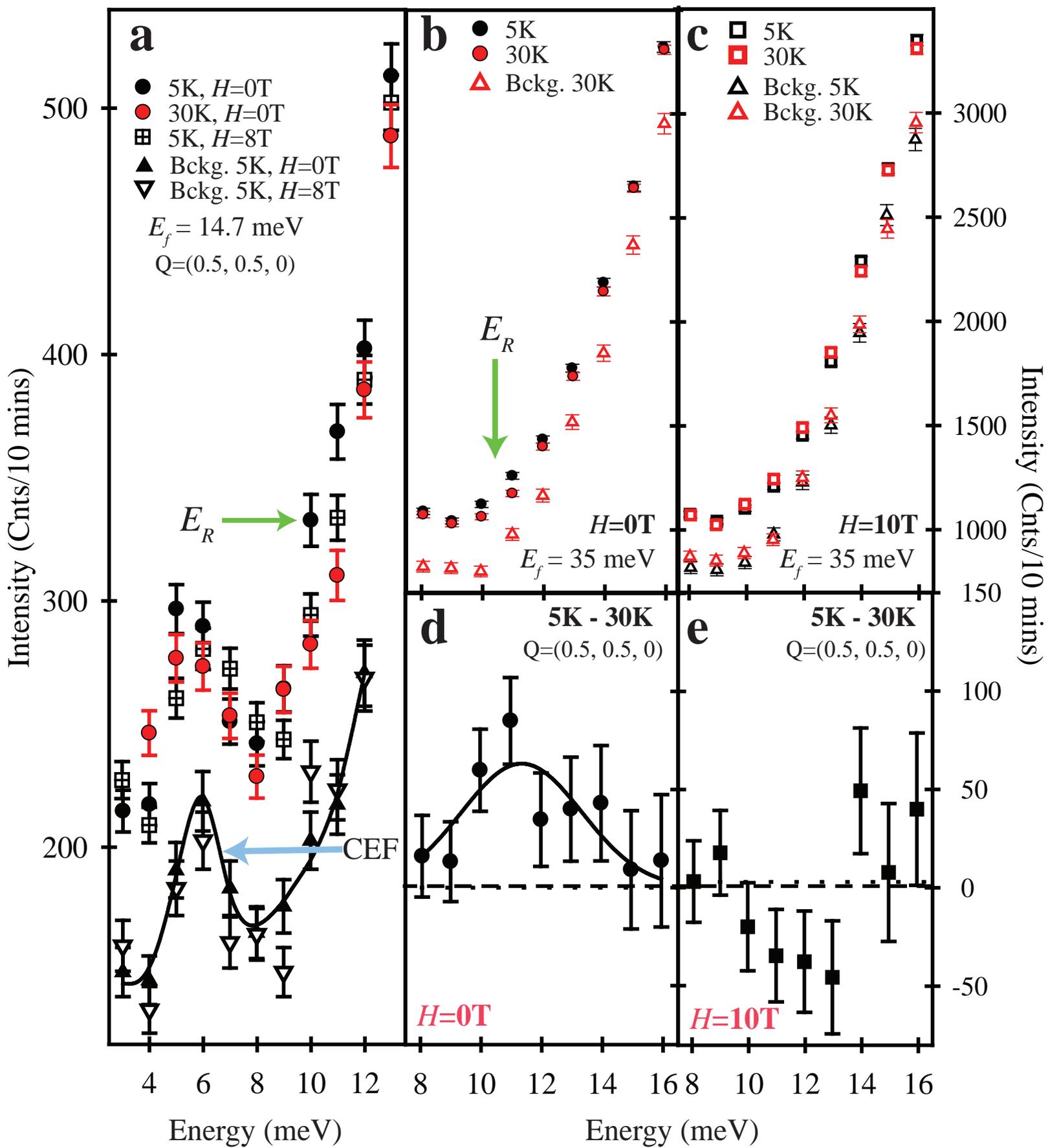

Figure 3

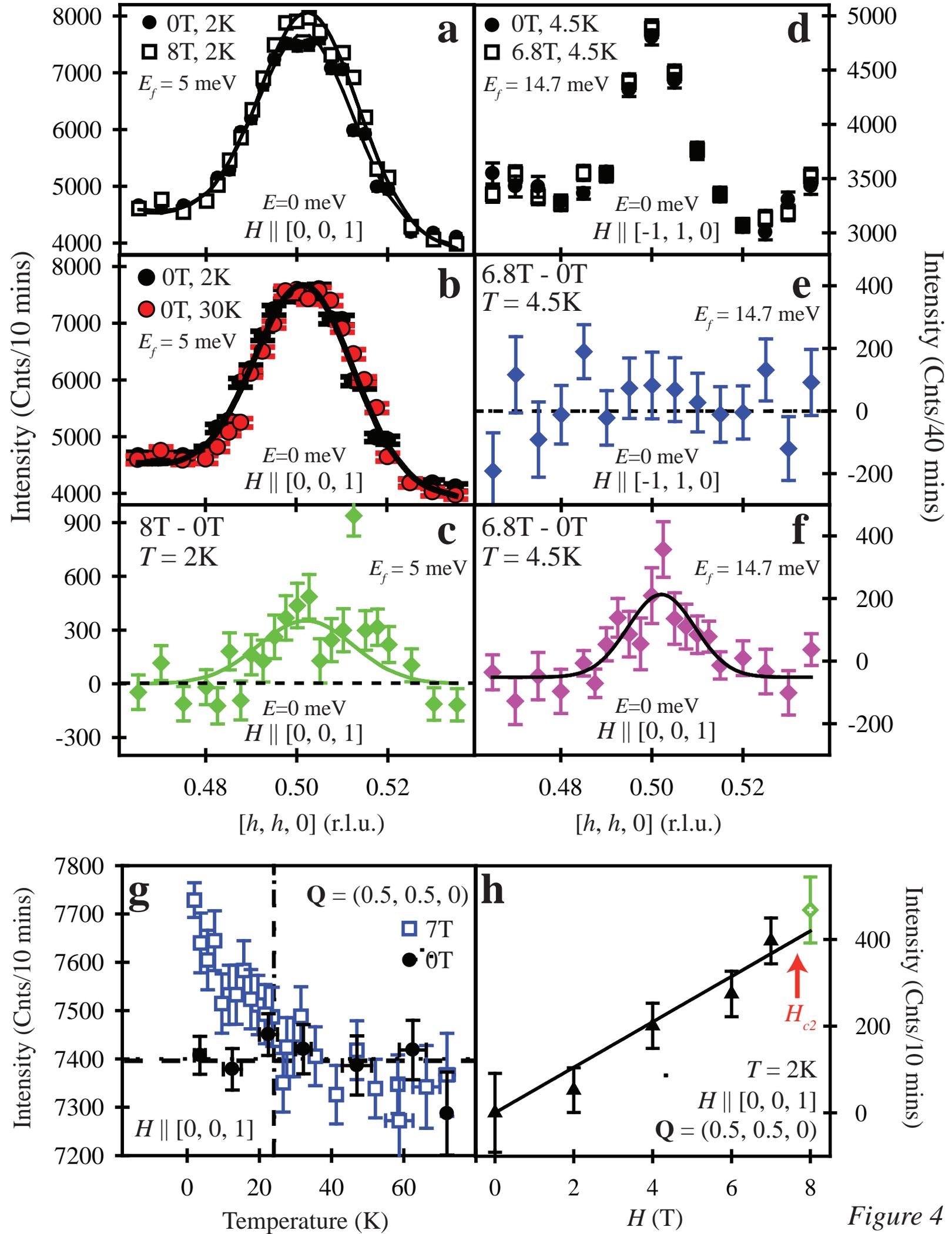

Figure 4

# Supporting Information

## 1. Specific Heat and Entropy Data

In order to expand upon the discussion of the specific heat measurements and resulting estimated condensation energy in PLCCO, we present detailed analysis of the entropy changes upon entering the superconducting phase in PLCCO($T_c$ = 24K). One measure of the importance of fluctuation effects in measurements of the specific heat in superconductors is to check the entropy balance condition for the quantity $(C/T)_{\text{Superconducting State}}$ -

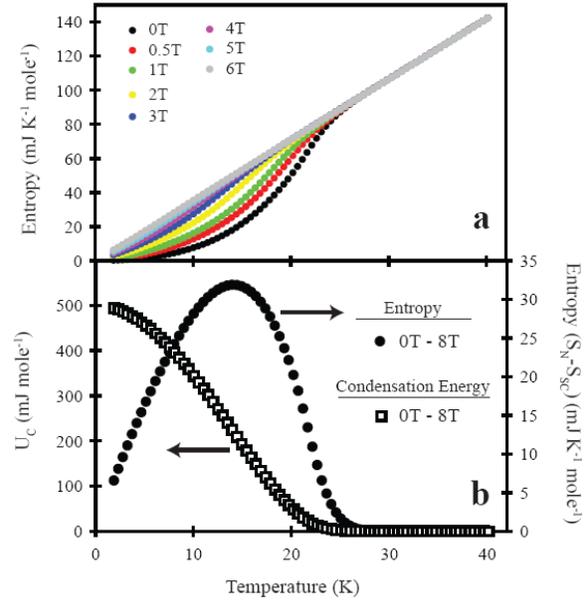

**SIFig. 1:** Entropy and condensation energy measurements for PLCCO ($T_c$ = 24 K). a) $S(T)$ measured under various field strengths and temperatures. b) Field subtracted $S(T)$ and condensation energy $U_c(T)$. The response of the nonsuperconducting field suppressed ground state at 8T was removed from zero field measurements in order to isolate entropy changes upon cooling through $T_c$.

$(C/T)_{\text{Normal State}}$ integrated up to $T_c$. Previous measurements on optimally electron-doped PCCO samples have shown that contrary to the case of hole-doped superconductors, such a balance is met at $T_c$ with no need to invoke the presence of a pseudogap contribution to the "normal state" electronic density of states[1]. In a similar vein, an analysis of the specific measurements presented in Fig. 1c of the main text confirms a similar balance condition met for the PLCCO ($T_c$ = 24 K) system. Performing an integration of $(C/T)_{\text{SC}}$-$(C/T)_{\text{NS}}$ for $T > 14$ K results in 31.8 mJ mole$^{-1}$ K$^{-1}$ for the positive area whereas integrating $(C/T)_{\text{SC}}$- $(C/T)_{\text{NS}}$ for $T < 14$K renders -32.8 mJ mole$^{-1}$ K$^{-1}$ for the negative area. The lowest-temperature $C/T$ value was linearly extrapolated to 0 K in order to approximate the entirety of the negative area. The 3% difference between these two values is well within the experimental uncertainty and the resulting entropy balance confirms the density of states present above $T_c$ and associated with the pseudogap in hole-doped high-$T_c$ systems is not present in this electron doped system. This further indicates that the potential influence of fluctuation effects on the estimate of the superconducting condensation energy used in our analysis is small.

Raw data showing the measured entropy $S(T)$ are plotted in SIFig. 1a for various field strengths. The decrease in the entropy at the onset of superconducting order ($T_c$) is evident in zero field and has completely vanished under the application of $H$~6 T. SIFig. 1b shows the field subtraction (0 T – 8 T) data for both $S(T)$ and the calculated condensation energy ($U_c$). Any small fluctuation effects still present in the field-suppressed normal state specific heat values are subtracted off through this procedure thereby removing their influence on the reported condensation energy for a given magnetic field strength. The fact that the entropy difference between the superconducting state and normal state in SIFig. 1b goes to zero at $T_c$ is further proof that fluctuation effects are minimal in this system. Superconducting fluctuations which are nonsingular yet contribute to the internal energy change as the system is cooled through $T_c$ however remain included in our calculation of the condensation energy; although it has been argued that even these potentially nonsingular portions of the internal energy are relevant in resolving the true condensation energy of the system[2].

## 2. Polarized neutron measurements of the resonance mode

In order to test the validity of the temperature subtraction method to isolate the magnetic scattering and verify that the resonance reported in the PLCCO ($T_c$ = 24 K) system[3] is genuine and not simply a sample specific property or nonmagnetic artifact, we performed polarized neutron measurements on a different set of optimally doped PLCCO ($T_c$ = 27 K) samples using IN22 at Institut Laue-Langevin (ILL), Grenoble, France. The results presented in the following section confirm that an identical resonance mode

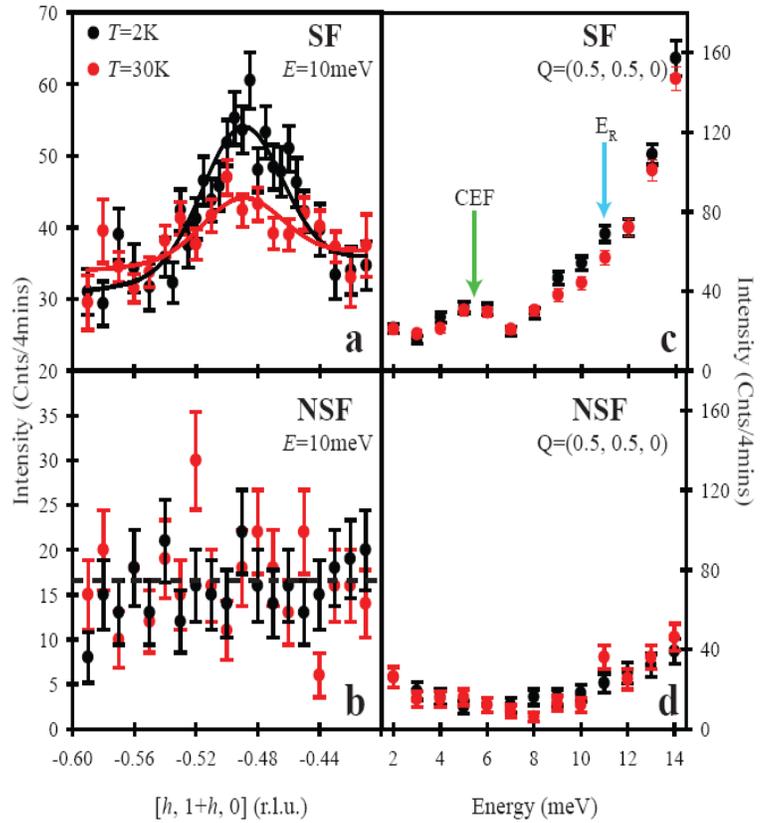

**SIFig. 2 :** Polarized neutron measurements of the resonance mode in PLCCO ($T_c$ = 27K). **a)** $\hbar\omega$ = 10 meV Q-scans at $T$ = 2 K and 30 K taken through (0.5, 0.5, 0). This data was collected in the spin-flip cross section channel. **b)** Identical measurements to those in panel *a* now instead collected in the non spin-flip cross section channel. **c)** Energy scans collected at Q = (0.5, 0.5, 0) at $T$ = 2 K and 30 K in the spin-flip scattering channel. The resonance mode and crystal electric field (CEF) excitations are denoted by arrows. **d)** Identical measurements to those in panel *c* collected instead in the non spin-flip scattering channel.

is present in this separate PLCCO system and that the experimental signature of the resonance given through a temperature difference in unpolarized neutron experiments is legitimate. The resonance therefore is a robust phenomenon within superconducting PLCCO and stands unquestionably as the electron-doped analog to the universally accepted resonance excitations observed within the hole-doped $YBa_2Cu_3O_{6+y}$ system.

For our experiments, we co-aligned three identical PLCCO ($T_c = 27$ K) samples within the [$H$, $K$, 0] scattering zone, and mounted them within a liquid He-cooled cryostat. Experiments were performed on the IN22 triple-axis spectrometer at ILL with a polarized spectrometer setup utilizing a cryopad to maintain zero guide field at the sample position[4]. Heusler-alloy monochromating and analyzing crystals were used in order to select the desired incident and final neutron polarizations. Within polarized neutron experiments, magnetic scattering can be isolated from nonmagnetic signal via the separation of the scattering cross sections into channels where the neutrons' incident spin direction has been flipped (spin-flip channel or SF) or not flipped (non spin-flip channel or NSF). In the following experiments, the neutron's polarization was tuned to be parallel to the scattering wave vector, $\mathbf{Q}$. Since the neutron is only sensitive to magnetic moments perpendicular to the scattering wave vector, this configuration effectively allows all resolvable magnetic scattering to be isolated within the spin-flip cross section channels[4].

Looking first at SIFigs. 2a-b, $\mathbf{Q}$-scans through $\mathbf{Q} = (0.5, 0.5, 0)$ at the resonance energy $E_R \sim 10$ meV show a correlated peak within the SF scattering channel centered at $\mathbf{Q} = (0.5, 0.5, 0)$. The same scans in the NSF channel show featureless scattering (SIFig. 2b), thus confirming the magnetic nature of the peak at $\mathbf{Q} = (0.5, 0.5, 0)$. There exists a clear enhancement in spectral weight at the $\mathbf{Q}=(0.5, 0.5, 0)$ position in the SF channel upon cooling into the SC phase (2 K) from the normal state (30 K). This is the hallmark of the resonance mode in this system and is consistent with our earlier unpolarized experimental work on the resonance mode in a PLCCO ($T_c = 24$ K) sample[3]. Looking at the NSF channel (which contains no magnetic contributions) in SIFig. 2b, no peak is observed at **Q**=(0.5, 0.5, 0) and only featureless, nonmagnetic background signal remains. Additionally, there is no change in the NSF channel upon cooling into the SC phase, further demonstrating that these nonmagnetic background processes have no influence on the reported resonance.

Energy scans taken at **Q**=(0.5, 0.5, 0) and plotted in SIFig. 2c-d also demonstrate that the resonance mode appears only within the SF channel. While the NSF channel shows no difference between 2 K and 30 K data, the SF channel shows a clear enhancement at the resonance energy upon cooling into the SC phase. The large intensity slope in the SF channel at higher energies in SIFig. 2c is due to magnetic scattering originating from the $Pr^{3+}$ CEF excitation at ~18 meV in this system[5].

The presence of the correlated peaks at **Q**=(0.5, 0.5, 0) exclusively in the SF channel definitively illustrates the magnetic origin of the resonance signal. Equally importantly, the enhancement of the spectral weight at **Q**=(0.5, 0.5, 0) at the resonance frequency upon cooling below $T_c$ appears only within the SF channel. This verifies that the entirety of the enhancement attributed to the resonance is magnetic and validates the use of direct temperature subtraction in previous unpolarized neutron measurements[3] (which yield identical results to those in SIFig. 2a ) and facilitates the temperature based separation of the resonance mode from background processes in the data presented within the main text of our paper here.

**3.** *Susceptibility measured with H parallel to the ab-plane and H parallel to the c-axis*

Although $Pr^{3+}$ in electron-doped PLCCO has a nonmagnetic singlet ground state[5,6], the pseudo-dipolar interaction between $Pr^{3+}$ and $Cu^{2+}$ can induce a small $Pr^{3+}$ moment in PLCCO. As a consequence, the low temperature the magnetic susceptibility of PLCCO is dominated by the $Pr^{3+}$ paramagnetic moment and is highly anisotropic with

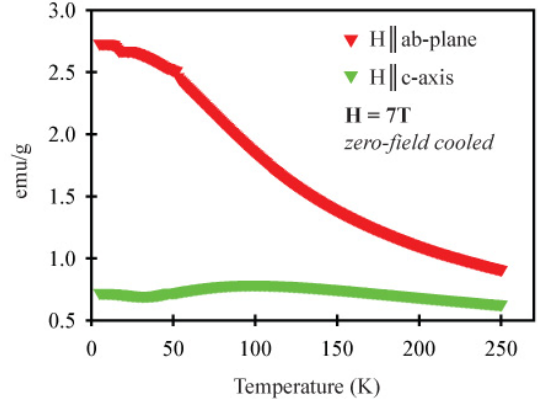

**SIFig. 3** The measured susceptibility for PLCCO for fields in the *ab*-plane (red) and along the *c*-axis (green). The data were collected on a SQUID magnetometer.

respect to the direction of an externally applied magnetic field. The susceptibility for a field along the moment easy axis direction in the $CuO_2$ plane is several times larger than for a perpendicular field (SIFig. 3). Therefore, one expects that it will be much easier to polarize the $Pr^{3+}$ moment for an in-plane magnetic field as compared to a *c*-axis aligned field, opposite to the effect of a field on superconductivity. Since experimentally we have observed a clear field-induced effect for a *c*-axis aligned field at **Q** = (0.5, 0.5, 0) but no effect for the same field in the $CuO_2$ plane, we conclude that the observed field-induced effect is likely to arise from the suppression of superconductivity.

### 4. *Field effect on the impurity phase*

In order to acquire a superconducting phase, electron-doped high-$T_c$ cuprates require a post-growth annealing process in an oxygen poor environment. This annealing process has been shown to create a small volume fraction (~1-2%) of a cubic impurity phase within the crystalline matrix of the cuprates[6-8]. For electron-doped

$Nd_{1.85}Ce_{0.15}CuO_4$ (NCCO), there have been much debate about whether the observed field-induced effect is due to NCCO or from the impurity phase[7,8]. In the case of PLCCO, this impurity phase is $(Pr_{1-x}LaCe_x)O_3$ whose cubic structure has a lattice constant match with the host lattice ($a_{impurity}=2\sqrt{2}$ a; a = 3.98 Å) that generates a nuclear Bragg reflection at the $\mathbf{Q} = (1/2, 1/2, 0)$ position[7,8]. Since this is one of the

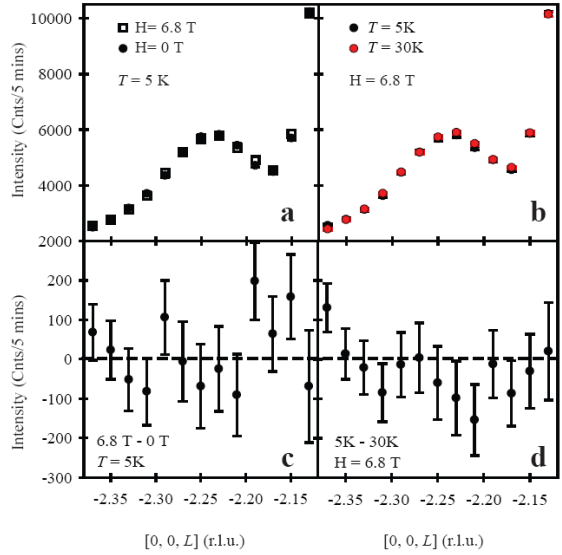

**SIFig. 4** The 6.8 T *ab*-plan field-induced effect on the cubic impurity phase $(Pr,La,Ce)_2O_3$ in PLCCO. The experimental geometry is the same as those in Fig. 4d and 4e. Consistent with ref. 6, a 6.8 T field has no observable effect on the impurity phase.

wave vectors where field-induced AF order appears in PLCCO, it is vital to differentiate between effects intrinsic to this impurity phase and the field induced effects originating from bulk superconducting phase of PLCCO. Fortunately, this is possible due to a *c*-axis lattice mismatch of ~10% between the imbedded impurity phase and the host lattice, where impurity reflections are also resolvable at (0, 0, 2.2), (0, 0, 4.4), … positions[6]. This allows any potential field effect on the impurity phase to be observed independently of the field induced AF order at the (1/2, 1/2, 0) position [6-8].

For our experiments, we studied the field dependence of the impurity $\mathbf{Q} = (0, 0, 2.2)$ reflections with the magnetic field applied parallel to the [-1, 1, 0] direction to minimize the effect on superconductivity (SIFigs. 4a-d). This allows for the direct observation of any potential field induced order from the polarization of the paramagnetic

$Pr^{3+}$ ions originating from the impurity phase exclusively. The outcome, shown in SIFig. 4, confirms that there is no field induced order originating from the $(Pr_{1-x}LaCe_x)O_3$ impurity phase in this PLCCO system for a 6.8 T field. Elastic **Q**-scans show no difference between scans taken in 0 T or 6.8 T in either the normal state (30 K) or the superconducting state (5 K). This observation confirms previous measurements by Kang *et al.* [6], in which no field effect on the impurity phase in PLCCO was observed over a series of doping levels. This further indicates that the observed *c*-axis field-induced effect is due to the suppression of superconductivity.

## 5. *Normalization of Resonance to Absolute Units*

It is of considerable interest to determine the absolute magnitude of the fluctuating moment involved with the resonance mode in PLCCO. To do this, we take advantage of the fact that upon cooling through $T_c$ in zero field, the only change in the spectral weight of $S(\mathbf{Q},\omega)$ in the spin excitations is the appearance of the resonance mode. In addition, we measured a low energy, transverse acoustic phonon and used the known scattering function to determine the spectrometer dependent constants. We then used this constant and measured the momentum and energy integrated cross scattering of the resonance mode to determine its magnitude in $\mu_B^2/Cu$.

The differential cross-section for coherent one phonon emission is[9,10]:

$$\frac{\partial^2 \sigma}{\partial \Omega \partial E} = \frac{k_f}{k_i} \frac{(2\pi)^3}{2v_0} \sum_{\tau,s} \frac{1}{\omega_s} \left| \sum_d \frac{\bar{b}_d}{\sqrt{M_d}} e^{-2W} e^{i\kappa \cdot d} (\kappa \cdot \vec{\xi}_{ds}) \right|^2 (n(\omega)+1) \partial(\omega - \omega_s) \partial(\kappa - q - \tau)$$

. This becomes for a phonon mode at a given ($\kappa$, $\omega$) simply:

$$\frac{\partial^2 \sigma}{\partial \Omega \partial E} = A \frac{\hbar^2 N}{2E(\vec{q})} \frac{k_f}{k_i} (n(\omega)+1) (\vec{\kappa} \cdot \vec{\xi}_{qs})^2 e^{U^2} |F(\tau)|^2 \partial(E - E(\vec{q})) \quad .$$

Here $\vec{\kappa} = \vec{\tau} + \vec{q}$ is the momentum transfer of the neutron, $N$ is the number of unit cells, $k_f$ and $k_i$ are the incident and final wavelengths of the neutron, $[n(\omega)+1)] = [1-\exp(\hbar\omega/k_B T)]^{-1}$, $\vec{\xi}_{qs}$ is the polarization vector of the phonon, $e^{U^2}$ is again the Debye-Waller factor, A is a spectrometer dependent constant, and $E(\mathbf{q})$ is the energy of the phonon mode. $F(\tau)$ is the weighted nuclear structure factor for the nuclear zone center wave vector $\vec{\tau}$ where

$$|F(\tau)|^2 = \left|\sum_d \frac{\bar{b}_d}{\sqrt{M_d}} e^{i\vec{\tau}\cdot\vec{d}}\right|^2$$ summed over sites in the unit cell. Approximating this cross section in the long wavelength limit, it becomes

$$\frac{\partial^2\sigma}{\partial\Omega\partial E} = A \frac{\hbar^2 N}{2E(\vec{q})} \frac{k_f}{k_i} (n(\omega)+1)(\vec{\kappa}\cdot\hat{\mathbf{e}}_{qs})^2 e^{-2W} \frac{1}{M} |G(\tau)|^2 \partial(E - E(\vec{q}))$$

where M is the mass of the unit cell, $\hat{\mathbf{e}}_{qs}$ is now instead a unit vector in the direction of atomic displacement for the phonon mode, and $G(\tau)$ is the nuclear structure factor[10]. This expression allows the spectrometer dependent constant, A, to then be determined through the measurement of a known phonon in the material.

The cross section for paramagnetic spin fluctuations is [10]:

$$\frac{\partial^2\sigma}{\partial\Omega\partial E} = A \frac{(\gamma r_o)^2}{4} \frac{k_f}{k_i} N |f(\vec{\kappa})|^2 e^{-2W} (n(\omega)+1) \frac{2}{\pi\mu_B^2} \chi''(\vec{\kappa},\omega)$$ where $f(\vec{\kappa})$ is the isotropic

magnetic form factor for $Cu^{2+}$. The local susceptibility $\chi''_{local}(\omega) = \frac{\int \chi''(Q,\omega)\partial^3 Q}{\int \partial^3 Q}$ was calculated at the (0.5, 0.5) in-plane wavevector via **Q**-scans performed at specific energies.

For our experiment, we measured a transverse acoustic phonon at **Q** = (0.12, 2, 0) and determined $A$ using the long-wavelength limit relation stated above. The raw data for the phonon intensity are shown with the resulting fit in SIFig. 5. Before calculating the integrated susceptibility, the background was removed through subtracting the measured nonmagnetic signal away from the correlated (0.5, 0.5) position as shown in Ref. 3. All data were corrected for λ/2 contamination in the monitor, and in our calculations for data at both 2 K and 30 K, the Debye-Waller factor was assumed to be 1. We then calculated the momentum integrated susceptibility at $\hbar\omega = 10$ meV in absolute units. In order to obtain a fuller picture of the dynamic susceptibility we cross-normalized the value for $\chi''_{local}$(10 meV) from data collected on IN-8 to previous data collected on HB-1 and reported in Ref. 3. Since data from this earlier experiment was not collected in a magnet, the increased signal intensities and relative intensity changes are easier to gauge along with the structure of $\chi''(\mathbf{Q}, \omega)$.

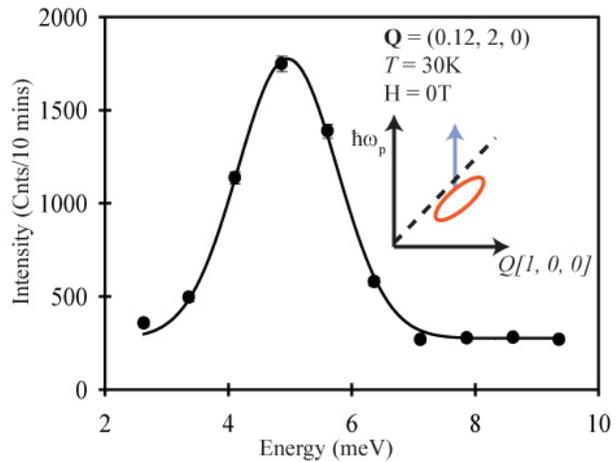

**SIFig. 5** The measured phonon scattering in IN-8 using the same setup as shown in Fig. 2a. One can estimate the absolute intensity of the resonance by comparing magnetic scattering with acoustic phonons.

For energies below 5 meV, the measured **Q**-widths along [$h$, $h$] were broader than the spectrometer resolution while scans at all higher energy transfers showed resolution limited widths along [$h$, $h$]. In estimating the local susceptibility, the magnetic signal was assumed to be a two-dimensional Gaussian within the [$h$, $k$] plane and rod-like out of the plane. This neglects the rotation of the resolution ellipsoid at energy transfers away from the resonance position and results in a slight underestimation of the integrated magnetic scattering at energies below the resonance. This estimation, however, is systematic and does not influence relative changes as a function of temperature in the local susceptibility as the system enters the superconducting phase.

Taking the difference between the local susceptibility determined at both 2 K and 30 K from 5 to 16 meV and integrating in energy, the resonance mode's total spectral weight can be determined. Performing this integration yields the value of $\langle m^2 \rangle = 0.0035 \pm 0.0015 \ \mu_B^2$ reported in the main text, with the large error bar resulting mainly from uncertainty in the energy width of the resonance mode.

Future measurements using a time of flight spectrometer to access the entire energy range from 0 meV to 100 meV in absolute units might allow us to accurately determine the zero-temperature *normal* state spin excitations.

## 6. Normalization of field-induce static AF Order to Absolute Units

Although it is in principle possible to also estimate the field-induced elastic scattering and compare the calculated moment to the suppression of the resonance at the inelastic position to test the total moment sum rule[11], such estimation requires the knowledge of the field-induced AF structure which is yet to be determined. In addition, it is difficult to compare measurements taken on a cold-triple spectrometer with those on a thermal triple-axis spectrometer because of the large differences in resolution volume and accessible reciprocal space volume. For these reasons, we have not attempted to convert our elastic measurements into absolute units.

## 7. Magnetic field-effect on low-energy spin fluctuations

In addition to the influence of a magnetic field on the resonance mode and on static AF order in PLCCO($T_c$ = 24 K), we also probed the response of the low energy magnetic fluctuations under the application of a c-axis aligned magnetic field. At zero field there exists a continuum of gapless, commensurate spin excitations down to $\hbar\omega$ = 0.5 meV [3,12], which have weak temperature dependence below 30 K with no direct coupling to $T_c$. Surprisingly, a clear suppression of the quasi-elastic magnetic scattering occurs under the influence of a magnetic field.

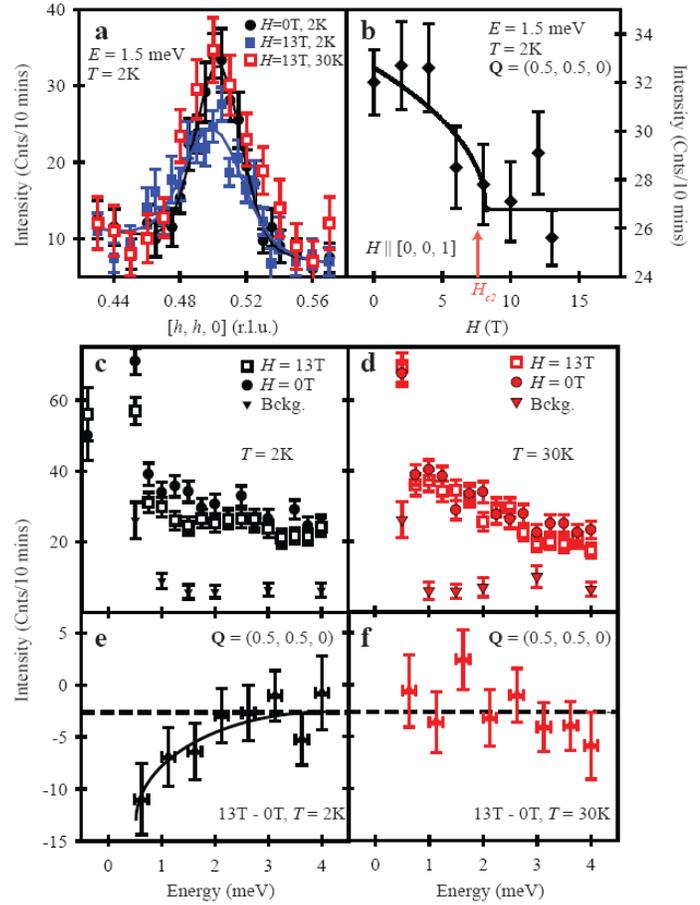

**SIFig. 6**: c-axis aligned magnetic field effect on the low energy spin excitations in PLCCO($T_c$=24K). **a)** $\hbar\omega$=1.5 meV Q-scans through (0.5, 0.5, 0) at 2K and 30K in 0T and 13T. **b)** H-dependence of spin fluctuations at $\hbar\omega$ = 1.5 meV and Q = (0.5, 0.5, 0) at low-T. **c-d)** $T$ = 2 K and $T$ = 30 K energy scans taken at both signal (Q=(0.5, 0.5, 0)) and background (Q=(0.6, 0.6, 0)) positions. Scans were collected both under zero field and under the application of H = 13 T. **e-f)** Field subtracted data from energy scans in panels c and d at both 2 K and 30 K respectively.

As shown in SIFigs. 6c and 6e, low temperature ($T$ = 2 K) energy scans at **Q** = (1/2, 1/2, 0) show a suppression of the low energy (0.5 meV < $\hbar\omega$ < 2 meV) magnetic scattering under 13 T. Upon warming into the normal state at $T$ = 30 K the spin fluctuations recover their intensity, and thus this field-induced suppression of the quasi-elastic scattering vanishes (SIFigs. 6d and 6f). The much weaker, energy-independent field-induced suppression shown in both the 2 K and

30 K field subtracted data (SIFigs. 6d and 6f, about 2 counts/10 minutes) is most likely due to a slight polarization of paramagnetic $Pr^{3+}$ moments.

The low-energy field-induced suppression at 2 K is shown more clearly through **Q**-scans around (0.5, 0.5, 0) at $\hbar\omega = 1.5$ meV (SIFig. 6a). SIFig. 3b suggests that the field-induced intensity reduction is associated with the suppression of superconductivity at $H_{c2}$. After warming to 30 K under 13 T, the intensity of the spin fluctuations at 1.5 meV merely recover their normal state 30 K value in 0 T, in violation to Bosonic statistics. Therefore, this spectral weight increase cannot simply arise from the population of spin waves emanating from the field-induced AF order but instead suggests a true recovery of the normal state scattering—a surprising finding since $T_c$ no longer remains a viable energy scale for $H \geq H_{c2}$.